\documentclass[pre,amsmath,amssymb,superscriptaddress,showpacs]{revtex4-1}
\usepackage{amsmath,amssymb,graphicx}
\usepackage{enumitem}
\usepackage{amsbsy}
\usepackage{latexsym}
\usepackage{color}
\usepackage{hyperref}
\usepackage{graphicx}
\usepackage{psfrag}
\usepackage[normalem]{ulem}
\usepackage{bm}
\usepackage{amsmath}
\usepackage{comment}

\newcommand{\be}{\begin{equation}}
\newcommand{\ee}{\end{equation}}
\newcommand{\bea}{\begin{eqnarray}}
\newcommand{\eea}{\end{eqnarray}}

\begin{document}

\title{Kinetics of the one-dimensional voter model with long-range interactions}

\author{Federico Corberi}
\email{fcorberi@unisa.it}
\affiliation{Dipartimento di Fisica ``E.~R. Caianiello'' and INFN, Gruppo Collegato di Salerno,
via Giovanni Paolo II 132, 84084 Fisciano (SA), Italy.}

\author{Claudio Castellano}
\email{claudio.castellano@roma1.infn.it}
\affiliation{Istituto dei Sistemi Complessi (ISC-CNR), Via dei Taurini 19, I-00185 Roma, Italy, and
Centro Ricerche Enrico Fermi, Piazza del Viminale, 1, I-00184 Rome, Italy.}

\begin{abstract}
The voter model is an extremely simple yet nontrivial prototypical model of ordering dynamics, which has been studied in great detail. Recently, a great
deal of activity has focused on long-range statistical physics models, where interactions take place among faraway sites, with a probability slowly decaying with distance. 
In this paper, we study analytically the one-dimensional long-range voter model, where an agent takes the opinion of another at distance $r$ with probability 
$\propto r^{-\alpha}$. The model displays rich and diverse features as $\alpha$ is changed. For $\alpha >3$ 
the behavior is similar to the one of the nearest-neighbor version, with the formation of ordered domains whose typical size grows as
$R(t)\propto t^{1/2}$ until consensus (a fully ordered configuration) is reached. 
The correlation function $C(r,t)$ between two agents at distance $r$ obeys dynamical scaling 
with sizeable corrections at large distances $r>r^*(t)$, slowly fading away in time.
For $2< \alpha \le 3$ violations of scaling appear, due to the simultaneous presence of two lengh-scales,
the size of domains growing as $t^{(\alpha-2)/(\alpha-1)}$, and the distance $L(t)\propto t^{1/(\alpha-1)}$ over which correlations extend. 
For $\alpha \le 2$ the system reaches a partially ordered stationary state, characterised by an algebraic correlator, 
whose lifetime diverges in the thermodynamic limit of infinitely many agents, so that consensus is not reached. 
For a finite system escape towards the fully ordered configuration is 
finally promoted by development of large distance correlations.
In a system of $N$ sites, global consensus is achieved after a time 
$T \propto N^2$ for $\alpha>3$, $T \propto N^{\alpha-1}$ for $2<\alpha \le 3$, and $T \propto N$ for $\alpha \le 2$.

\end{abstract}

\maketitle

\section{introduction}\label{intro}

Understanding the evolution of systems far from equilibrium is a central issue of 
modern physics and beyond. 
In this framework, phase-ordering -- the process whereby a multiphase system demixes -- is perhaps
the most studied and better understood phenomenon, with applications ranging  from alloys, fluids and other physical systems ~\cite{Bray94,PuriWad09,CRPHYS_2015__16_3_257_0} to population dynamics~\cite{10.2307/j.ctv36zpzm}. With short-range interactions, the ordering
process is characterised by the increase of a typical 
length $R(t)$, associated to the size of demixed regions, known as coarsening. A notable example is the Ising model with Glauber kinetics quenched to low temperatures, where $R(t)\propto t^{1/2}$.  This growth phenomenon is usually characterized by a dynamical scaling symmetry that, in a nutshell, means that $R(t)$ is the only relevant length scale present in the system and that the whole
effect of elapsing time is the mere increase of such length, leaving all other features of the system unchanged. 
Stated differently, two configurations at different times look statistically similar if lengths are
measured in units of $R(t)$. This particular symmetry is reflected in the behavior of physical observables
among which the equal-time correlation function. For a binary system described by boolean variables,
or {\it spins}, $\{S_i\}=\{\pm 1\}$, defined on sites $i=1,\dots,N$ of a $d$-dimensional lattice, this is defined as
 $C(r,t) =\langle S_i(t)S_{j}(t)\rangle$, where $r$ is the $i-j$ distance. 
When phase-ordering occurs well away from 
a critical point, the above symmetry, implies, for sufficiently long times, 
\be
C(r,t)=f\left (\frac{r}{R(t)}\right ),
\label{scalAnsatzSimple}
\ee
where $f(x)$ is a scaling function with limits $f(x \to 0)=1$ and $f(x \to \infty)=0$. 

Despite the rather good level of comprehension of these processes in systems with short-range interactions, less in known when they are long-range~\cite{Bray94,CLP_review,CMJ19,iannone2021, CLP_lambda, CLP_epl, PhysRevE.103.012108, BrayRut94, RutBray94}, because computer simulations are numerically demanding and analytical approaches are scarce. In particular,
considering the paradigmatic case where the coupling between spins decays 
algebraically as $r^{-\alpha}$, a first important distinction must be made between the so called
{\it weak long-range} case~\cite{Defenu_2020}, when $\alpha >d$ and the interaction is integrable, and the 
{\it strong long-range} situation with $\alpha \le d$. In the former case the behavior of the system is 
generally similar to that of the the corresponding short-range case, and an analogous pattern
characterised by growing domains is found, associated to the dynamical scaling symmetry.
At a more quantitative level, there exists a characteristic value $\alpha_{SR}>d$ above which
scaling exponents, such as the one regulating the growth of the length $R(t)\propto t^{1/z}$,
keep the same value of the corresponding model with nearest neighbor (NN) interactions, 
whereas they change and become $\alpha$-dependent below $\alpha_{SR}$. 
In the $1d$ Ising model, for instance, it is $\alpha_{SR}=2$~\cite{CLP_review,CLP_epl,BrayRut94,RutBray94}.
The situation changes dramatically when entering the strong long-range regime with $\alpha <d$, because in this case 
some of the very fundaments of statistical mechanics, notably additivity, break down and a 
physically very different behavior is expected~\cite{campa2009statistical, book_long_range, DauRufAriWilk}.
It must be observed, indeed, that the extreme limit $\alpha \to 0$ corresponds to the mean-field
case where, by definition, domains cannot exist and equilibration occurs through a completely
different mechanism, without any coarsening. There must exist, therefore, a radical change
of behavior when $\alpha $ is lowered. However, while for the equilibrium critical properties such a 
crossover to a mean-field like behavior is known to take place upon crossing a well defined value of $\alpha$,
for the dynamical properties the situation is possibly less simple and much less is known~\cite{iannone2021}.

In this paper we give a contribution by solving analytically the voter model~\cite{liggett2013stochastic} with algebraic interactions in one dimension. 
Voter dynamics is defined in terms of the same spin variables of the Ising model with the
difference that a spin $S_i$ takes the value of another $S_j$, chosen at random among its nearest neighbors.
At variance with the corresponding Ising model, this rule does not obey detailed balance and, hence, 
it is not possible to define an Hamiltonian, except for the one-dimensional case when the voter and the Ising model at $T=0$ can be mapped onto each other. 
The voter model with nearest-neighbor interactions is an extremely interesting model for several reasons: it is one the few nontrivial out of equilibrium dynamics that can be solved exactly in any dimension~\cite{Frachebourg96}; it undergoes a coarsening process in the absence of surface 
tension, thus exhibiting nontrivial critical properties~\cite{Dornic01, AlHammal05, Castellano12}; in social dynamics it is a paradigmatic model for the formation of consensus~\cite{Castellano09,Redner19}.
Here we consider a long-range one-dimensional voter model, with spin $S_i$ taking the value of another $S_j$, chosen at distance $r$ with probability $P(r)\propto r^{-\alpha}$.
Considering the process starting from a disordered configuration, 
our study enlightens a rich pattern of different behaviors, where the
scaling symmetry takes different forms and is characterised by different exponents as $\alpha$ is changed,
or is violated in different ways. The transition to the mean-field regime is, itself, interesting and unconventional.   
The one-dimensional voter model with long-range interactions had been introduced and studied numerically before~\cite{PhysRevE.83.011110}. Our analytical results confirm
and complete some of the findings of that work, while correcting some conclusions which were extrapolated from simulations.

Before presenting, in the next sections, the analytical solution of the model, we summarise below the main results,
discussing also how they are organised into the various parts of this article.
We start, in the next section~\ref{secmodel}, by defining the model, discussing the main observables we will consider, and 
deriving the evolution equation for $C(r,t)$, in terms of which the problem can be analytically studied.
Sec.~\ref{agt3} is devoted to the solution of the model in the case $\alpha >3$. 
It is found that $\alpha_{SR}=3$ and, hence, for $\alpha >3$
the behavior is in the same universality of the short-range model. Indeed one finds that in this
case the growth law is $R(t)\propto t^{1/2}$.
One also observes the scaling form~(\ref{scalAnsatzSimple}), with $f(x)$ a monotonically decreasing function
of exponential type; however this is true only up to $x\propto x^*(t)=\sqrt{(\alpha -3)\ln t}$. Beyond this value Eq.~(\ref{scalAnsatzSimple}) is yet obeyed but with respect to another length $\mathcal{L}(t)\propto t^{3/(2\alpha)}$,
and with an algebraic decay $f(x)\propto x^{-\alpha}$ of the scaling function.
This behavior gradually transforms into the standard one as time elapses, because $x^*(t)$ increases 
and the large-$x$ form is pushed to regions where correlations are negligible. However, note that this process
is logarithmically slow and further depressed as $\alpha \to 3$. 
In the following Sec.~\ref{ain23} the solution of the model is derived for $2<\alpha \le3$.
In this case one has a violation of scaling: two distinct growing lengths are present. This occurs because of the behavior of the correlation function at $r=1$. Let us remind that $C(r=1,t)$ is 
directly related to the density of interfaces $\rho (t)$ in the system (adjacent anti-aligned spins) and that $\rho^{-1}$ is the size of domains. 
In this range of $\alpha$-values one finds $\rho^{-1}(t)\propto t^{(\alpha -2)/(\alpha -1)}$ whereas the size of correlated regions grows $\propto t^{1/(\alpha -1)}$. Note that now, at variance with the previous case with $\alpha >3$, scaling is not progressively
recovered as time goes by.
Finally, the case with $0<\alpha \le 2$ is discussed in Sec.~\ref{aless2}.
The new feature, addressed in Subsec.~\ref{ninftyfirst}, is now that in the infinite-size limit a stationary state is reached,
characterised by lack of consensus, and exhibiting  correlations extending
to very long distances, as testified by the slow algebraic decay of $C(r)\propto r^{-(2-\alpha)}$, for $1<\alpha \le 2$, or $C(r)\propto r^{-\alpha}$ for $0\le \alpha\le 1$. Similar non-equilibrium stationary states are commonly observed in systems with strong long-range interactions.
The stationary state is preceded by a coarsening stage, as discussed in Subsec.~\ref{approachstat}.
Sec.~\ref{ndependence} contains a discussion of the dependence of physical quantities on the size $N$ of the system.
Indeed, due to the long-range nature of the interaction, for $\alpha \le 2$ even intensive quantities turn out to depend on $N$,
a fact that has important consequences, for instance, on the time to reach consensus, which is studied in Sec.~\ref{secconsensus}.
The last section, Sec.~\ref{concl}, contains a summary of the results obtained, some comparison with other
coarsening models, and a perspective for future investigations.

\section{The model and its dynamical equation} \label{secmodel}

The voter model is defined in terms of boolean (or spin) variables $S_i=\pm 1$, where $i=1,\dots,N$
runs over the sites of a one-dimensional lattice with periodic boundary conditions. In an elementary move a randomly chosen variable $S_i$ takes the state of another, $S_j$, 
chosen at distance $r$ with probability
$P(r)=\frac{1}{2Z}\,r^{-\alpha}$, where $Z=\sum_{r=1}^{N/2}r^{-\alpha}$. 
The probability to flip $S_i$ in a single update can than be written as
\be
w(S_i)=\frac{1}{2N}\sum _{r=1}^{N/2}P(r)\sum _{k=[[i\pm r]]}(1-S_iS_k),
\label{transrates}
\ee
where the factor $1/N$ describes the probability to randomly pick up $S_i$.
The symbol $[[\dots]]$
means that the boundary conditions must
be correctly taken into account, so, for instance, $[[i+r]]=i+r-N$ if $i+r>N$.
For an $n$-spin correlator it is easy~\cite{Glauber} to obtain
$\frac{d}{dt}\langle S_{i_1}S_{i_2}\cdots S_{i_n}\rangle=-2\langle S_{i_1}S_{i_2}\cdots S_{i_n}\cdot \sum _{m=1}^n w_\alpha(S_{i_m})\rangle$
which, specified for $n=2$, gives
\be
N\frac{d}{dt}\langle S_i(t)S_j(t)\rangle=-2\langle S_i(t)S_j(t)\rangle+\sum _{r=1}^{N/2}P(r)\left [\sum_{k=[[i\pm r]]}\langle S_j(t)S_k(t)\rangle
+\sum _{q=[[j\pm r]]}\langle S_i(t)S_q(t)\rangle \right ].
\label{eqc1}
\ee
In the following, as customary, we measure time in units of $N$, namely in Monte Carlo steps per spin, hence
$N\frac {d}{dt} \to \frac{d}{dt}$ on the l.h.s. of the previous equation and, assuming space translation invariance, one arrives at a closed equation
for the correlation function
\be
\frac{d}{dt}C(r,t)=-2C(r,t)+2\sum_{\ell=1}^{N/2}P(\ell)\left [C([[ r-\ell]],t)+C([[r+\ell]],t)\right ],
\label{eqc2}
\ee
where, due to the chosen boundary conditions,
\be
[[n]]=\left \{ \begin{array}{lll}
|n|, & \mbox{if}&  |n|\le N/2 \\
N-|n|, & \mbox{if} &|n|>N/2. 
\end{array} \right .
\ee 
From $C(r,t)$ a correlation length can be extracted as
\be
L(t)=\frac{\sum _{r=0}^{N/2}r\,C(r,t)}{\sum _{r=0}^{N/2}C(r,t)}.
\label{eqL}
\ee
Another length scale can be naturally
introduced. The density of domain walls (anti-aligned adjacent spins) 
is given by
\be
\rho(t)=\frac{1}{2}[1-C(r=1,t)].
\label{defrho}
\ee
The characteristic size of the domains is given by $\rho(t)^{-1}$.
When scaling holds it must be $\rho(t)^{-1}\propto L(t)$; 
however, as we will see, violations of dynamical scaling may occur,
leading to different evolutions of the two length scales.

Eq.~(\ref{eqc2}) is exact. Based on it we will determine in the following sections the behavior of $C$, starting from a fully disordered initial state, considering separately the various ranges of $\alpha$ values.

\section{Case \boldmath{$\alpha >3$}} \label{agt3}

\begin{figure}[ht]
	\vspace{1.0cm}
	\centering
 \rotatebox{0}{\resizebox{0.5\textwidth}{!}{\includegraphics{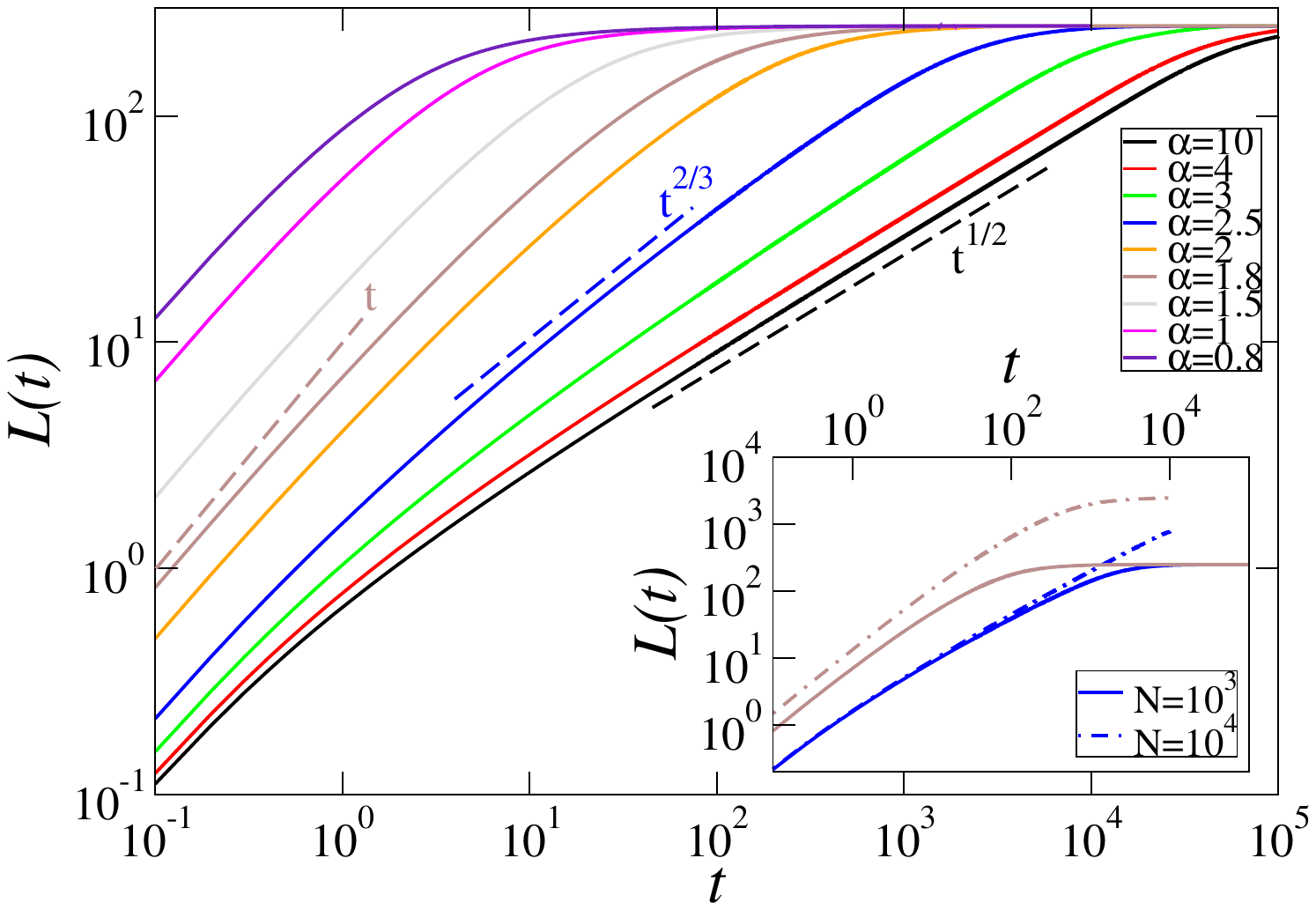}}} 
  \caption{$L(t)$ is plotted against time $t$, for different values of $\alpha$ (see legend). Data are obtained by solving numerically Eq.~(\ref{eqc2}) with $N=10^3$. 
  Dashed lines (with corresponding color as numerical data) are the analytical results of Eqs.~(\ref{diffgrowth},\ref{alphareg},\ref{balreg}). In the 
  inset a comparison is shown, for $\alpha=2.5$ and $\alpha=1.8$, with a larger system with $N=10^4$ (dot-dashed lines). Note that all curves saturate to $N/4$, which corresponds,  according to Eq.~\eqref{eqL}, to the fully ordered state.}
	\label{figL}
\end{figure}

For large enough $\alpha$ 
 it seems natural to search for a solution with properties similar to
the NN case. Recalling that for NN the scaling property holds~\cite{Glauber}, we plug the ansatz $C(r,t) = f[r/L(t)]$ into Eq.~(\ref{eqc2}), obtaining
\be
-xf'(x)\frac{\dot L(t)}{L(t)}=-2f(x)+2\sum_{\ell=1}^{N/2}P(\ell) \left \{f(|x-y|)+f([[x+y]])\right \}
\label{eqexp}
\ee
with $x=\frac{r}{L(t)}$ and $y=\frac{\ell}{L(t)}$. 
Let us now focus on the small-$x$ region. 
Assuming that $f(x)$ decays fast, the dominating contributions to the sum on the r.h.s. of Eq.~(\ref{eqexp}) are provided by the values 
$y\simeq x$ and $y\simeq 0$, in the first and second term, respectively. 
Considering $x$ small, the condition $y\simeq x$ implies that $y$ is small as well.
Hence we can Taylor-expand the functions on the r.h.s. of Eq.~(\ref{eqexp}) to second order in $y$,
thus obtaining
\be
-xf'(x)\frac{\dot L(t)}{L(t)}=\frac{2f''(x) \langle \ell^2 \rangle}{L(t)^2},
\label{eqqsscal}
\ee
where
\be
\langle \ell^2 \rangle = \sum_{\ell=1}^{N/2}\ell ^2 P(\ell)
\ee
is the second moment of the distribution of interaction distances.
Note that Eq.~\eqref{eqqsscal} is valid only for $\alpha>3$, otherwise
$\langle \ell^2 \rangle$ diverges. 
In order for any explicit time-dependence to cancel out in Eq.~(\ref{eqqsscal}) one must have 
\be
L(t)=D\,t^{1/2},
\label{diffgrowth}
\ee
where $D$ is a constant. This behavior is very clearly observed in Fig.~\ref{figL} where this quantity has been plotted after solving numerically Eq.~(\ref{eqc2}). After an initial transient, the curves for $\alpha >3$ show $L(t)$ growing as in Eq.~(\ref{diffgrowth}) before saturating when $L(t)$ approaches $N$,
when the finite size is felt and the system reaches consensus, namely a fully
ordered state (see Sec.~\ref{secconsensus} on this topic).

The scaling function must then obey the following equation
\be
\frac{x_0^2}{2} f''(x)+xf'(x)=0,
\ee
with 
$x_0=\frac{2\langle \ell^2 \rangle^{1/2}}{D}$.
The solution is 
\be
f(x)=c_1 \, \mbox{erf} \left (\frac{x}{x_0}\right )+c_2,
\ee
where $c_1,c_2$ are constants and $\mbox{erf}(x)=\frac{2}{\sqrt \pi}\int _0^x e^{-t^2}\,dt$. Fixing the constants using 
$\lim _{x\to 0}f(x)=1$ and $\lim_{x\to \infty}f(x)=0$, one finds
\be
f(x)=\mbox{erfc} \left (\frac{x}{x_0} \right ),
\label{psiafter3}
\ee
where $\mbox{erfc}(x)=1-\mbox{erf}(x)$.
Let us remark that all the above is valid for small $x$ and hence also for $r=1$. Then, recalling the definition~(\ref{defrho}), one finds $\rho(t)\propto L(t)^{-1}$.
A further comment is in order: we have used above the boundary condition
$\lim _{x\to \infty}f(x)=0$ despite the fact that the present solution is valid only for small $x$. This is however justified because the validity of the solution extends up to a certain $x^*(t)$ which, as shown below, diverges with time (see Eq.~(\ref{xstar})). Hence the boundary condition becomes correct for increasingly larger times.

Let us now investigate the large-$x$ behavior. The function $C([[r-\ell]],t)$ in Eq.~(\ref{eqc2}) has a peak of unitary height for $\ell =r$.
It can be verified for consistency at the end of the calculation that for large $x$ 
such a peak dominates the sum over $\ell$ and the entire r.h.s. in 
Eq.~(\ref {eqc2}).
Expanding $f(x)$ (in Eq.~(\ref{psiafter3})) to leading order to evaluate this
term near $\ell=r$, $f(x)\simeq 1-2/\sqrt \pi\, (x/x_0)$,
and approximating in this way the sum in Eq.~(\ref{eqc2}), one arrives at
\be
\dot C(r,t)\propto x_0 P (r) L(t),
\ee
leading to
\be
C(r,t) = \left ( a \,\frac{r}{
\mathcal{L}(t)}\right )^{-\alpha},
\label{xmenalfa}
\ee
where $a$ is a constant and $\mathcal{L}(t)\propto x_0^{1/\alpha}t^{3/(2\alpha)}$. This is again a scaling form of the type~(\ref{scalAnsatzSimple})
but with respect to a different length scale $\mathcal{L}(t)$ (in place of $L(t)$) and with a different, algebraic, scaling function.
Hence, for large but finite times, the scaling pattern given by Eq.~\eqref{psiafter3} extends
only up to a time-dependent value $x^*(t)=r^*/L(t)$, after which the correlation function crosses over from an exponential to an algebraic decay.
The crossover point can be found by matching the two scaling forms~(\ref{psiafter3},\ref{xmenalfa}) and using the 
behavior of the erfc function for large argument, erfc$(x)\simeq \frac{e^{-x^2}}{x\sqrt \pi}$, finding
\be
x^*(t)\propto \sqrt{(\alpha -3)\ln t}.
\label{xstar}
\ee

 \begin{figure}[ht]
	\vspace{1.0cm}
	\centering
 \rotatebox{0}{\resizebox{0.5\textwidth}{!}{\includegraphics{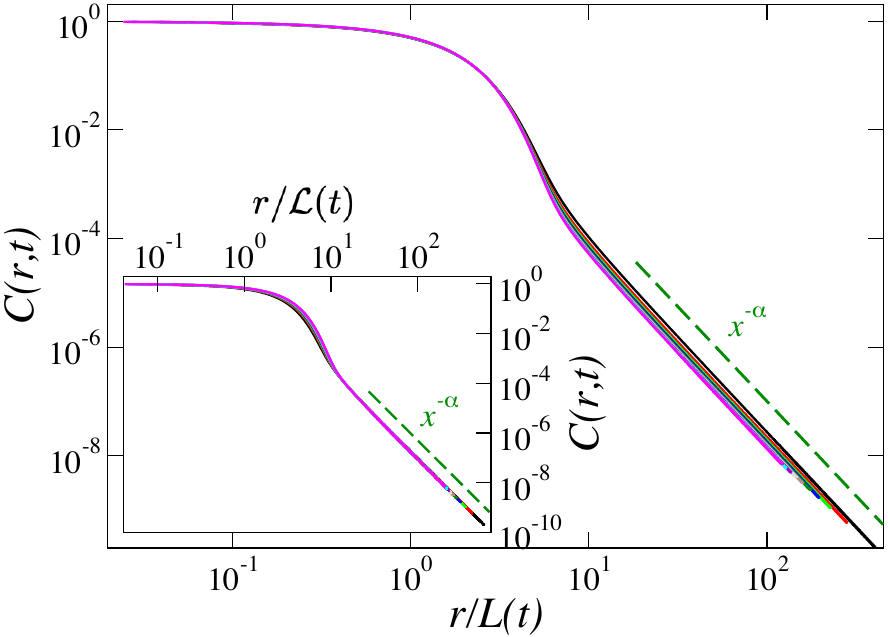}}} 
  \caption{Collapse plot, for $\alpha=3.5$, of the correlation function $C(r,t)$ as a function of $x=r/L(t)$ (main plot), or $x=r/\mathcal{L}(t)$ (inset). $L(t)$ is computed using Eq.~\eqref{eqL} and $\mathcal{L}(t)$ as defined below Eq.~(\ref{xmenalfa}). System size is $N=10^4$. The curves are for values of $t$ ranging from $10^2$ to $10^3$ in steps of $10^2$.}
	\label{fig_C1}
\end{figure}

As time elapses the preasymptotic large-distance behavior is pushed to larger and larger distances and, at the same time, to smaller and smaller values of the correlation. 
But this process is logarithmically slow and, moreover is less and less effective as $\alpha \to 3$, where a different solution, discussed below, comes into play.
The main part of Fig.~\ref{fig_C1} confirms that scaling holds for a slowly increasing
interval of $x$ values. For larger $x$ the decay is power-law as $x^{-\alpha}$, according to Eq.~(\ref{xmenalfa}), 
but curves
for different times do not perfectly fall onto each other.
This is because for $x>x^*$ scaling occurs with respect to 
the different length $\mathcal{L}(t)$ (defined below Eq.~(\ref{xmenalfa})) and, indeed, using this scaling length 
in the inset of the figure one finds an excellent collapse for large $x$ (but not for small $x$).

\section{Case \boldmath{$2<\alpha \le 3$}}\label{ain23}

As noted before, the behavior found in the previous section  breaks down for $\alpha \le 3$, as $\langle \ell^2 \rangle$ diverges.
We now show that for $2< \alpha \le 3$ the correlation function $C(r,t)$ for $r\gg 1$ obeys a scaling form analogous to the one in Eq.~(\ref{xmenalfa}), namely 
\be
C(r,t) \simeq f(x) = (ax)^{-\alpha},
\label{unomenxbis}
\ee
with $x=r/L(t)$,
but with a scaling length $L(t)$ growing differently from $\mathcal L(t)$ (see Eq.~(\ref{alphareg})). 

As before, also in this case the term containing
$C([[r-\ell]],t)$ provides the dominant contribution to the r.h.s. of Eq.~(\ref{eqc2}) for $r\gg 1$
(it can be checked for consistency at the end of the calculation).
Writing the sum as an integral, such a term, denoted as $S(x,t)$, reads
\be
S(x,t)\simeq 2L(t)\int_{1/L(t)}^{N/2L(t)}dy\, P[yL(t)] C(|x-y|L(t),t),
\label{eqS}
\ee
with $y=\ell /L(t)$.

The integral is dominated by the region around the peak of the integrand at $y \approx x$, whose height is $P[xL(t)]C(0,t)=L(t)^{-\alpha}(2Z)^{-1}x^{-\alpha}$ and whose width, according to Eq.~(\ref{unomenxbis}), is of order $a^{-1}$. Hence
\be
S(x,t)\simeq   
\frac{L(t)^{1-\alpha}x^{-\alpha}}{a Z}.
\label{I0}
\ee

Then, still considering large $r$, we can insert the form~(\ref{unomenxbis}) into the l.h.s. of Eq.~(\ref{eqexp}), and 
evaluate the r.h.s. by the expression~(\ref{I0}). 
The $x$-dependence of both the l.h.s. and the r.h.s. is the same. 
Imposing any explicit time-dependence to cancel out, the growth-law
\be
L(t)=B t^{1/(\alpha-1)}
\label{alphareg}
\ee
is obtained, with 
\be
B=a\left [\frac{2(\alpha -1)}{\alpha Z}\right ]^{1/(\alpha -1)}.
\label{eqBalpha}
\ee

The behavior~(\ref{alphareg}) is well reproduced by the numerical solution of Eq.~(\ref{eqc2}), as shown in Fig.~\ref{figL} for the case with $\alpha=2.5$.

\begin{figure}[ht]
	\vspace{1.0cm}
	\centering
 \rotatebox{0}{\resizebox{0.5\textwidth}{!}{\includegraphics{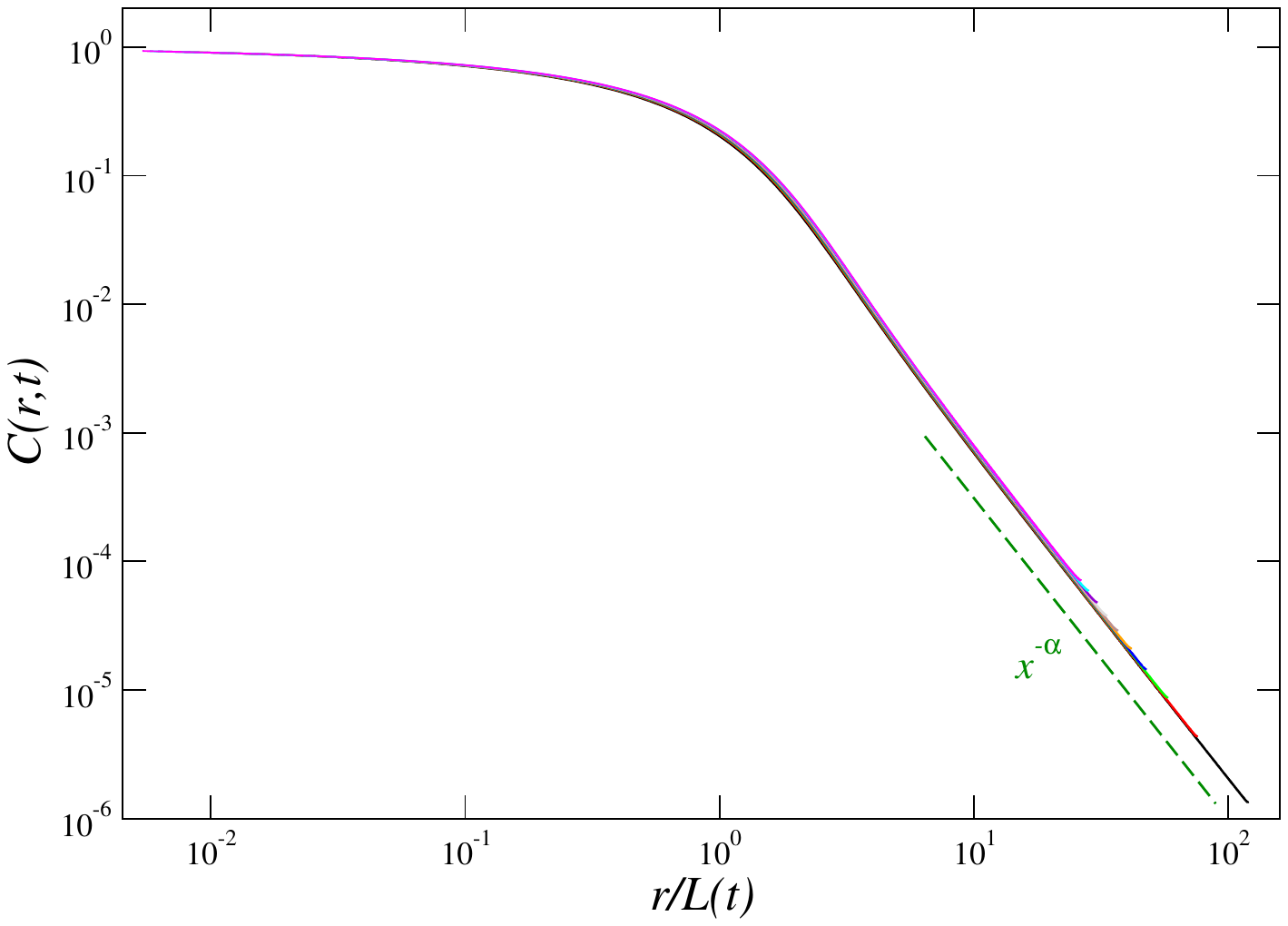}}}
  \caption{Collapse plot, for $\alpha=2.5$, of the correlation function $C(r,t)$ as a function of $x=r/L(t)$ where $L(t)$ is computed using Eq.~\eqref{eqL}. System size is $N=10^4$. The curves are for values of $t$ ranging (in steps of $10^2$ time units) from $10^2$ to $10^3$. The green dashed line is the behavior~(\ref{unomenxbis}). 
 }
	\label{fig_C2}
\end{figure}

In addition, Fig.~\ref{fig_C2} confirms that scaling holds and that the
scaling function decreases as Eq.~\eqref{unomenxbis} for large $x$.

Now we turn our attention to the value at $r=1$ where, clearly, Eq.~\eqref{unomenxbis} cannot hold asymptotically. We recall that the density of interfaces (anti-aligned adjacent spins) $\rho$ is related to $C(r=1,t)$ by $C(r=1,t)=1-2\rho$.
In Appendix~\ref{appdensint} it is shown that
\be
\rho(t) \sim \frac{k'}{2} L(t)^{-\beta},
\ee
where $k'$ is a constant.
Hence the size of domains $\rho^{-1}$ and the length  $L$ grow differently and scaling is violated. 
Unfortunately, the analytical determination of $\beta $ turns out to
be very difficult,
therefore we have carefully investigated this point numerically. The results of this study, presented in Appendix~\ref{appbeta}, provide evidence that $\beta=\alpha-2$. (Consistently, for this
value of $\beta$ we find $k'=(-\alpha^2+4\alpha-3)/(5-\alpha)>0$).
This implies 
\be
\rho^{-1}(t) \propto t^{(\alpha-2)/(\alpha-1)}.
\label{expdecayrhot}
\ee
Note that the growth exponent matches the short-range value $1/2$ as $\alpha \to 3$. 
Conversely, the growth exponent vanishes as $\alpha \to 2$, signalling
that, as shown in the next section, for $\alpha \le 2$ the system reaches
a partially-ordered non-equilibrium stationary state.

\section{Case \boldmath{$0 \le \alpha \le 2$}}\label{aless2}

For $\alpha \le 2$ in the thermodynamic limit $N\to \infty$ consensus is not reached (see below), while it is reached for any finite $N$ if we take the $t \to \infty$ limit first: the two limits do not commute.

\subsection{Stationary state} \label{ninftyfirst}

When the thermodynamic limit is taken first, for $1< \alpha \le 2$,
a non-trivial stationary state is attained, with 
\be
C_{stat}(r)=\kappa \,r^{-(2-\alpha)},
\label{Cstat}
\ee
for $r\gg 1$, where
$\kappa$ is a constant, see Appendix~\ref{ain12}. 
Note that, using the expression~(\ref{Cstat}) in Eq.~(\ref{eqL}) and transforming the sum into integrals one easily obtains the correlation length in the stationary state as
\be
L_{stat}=\frac{\alpha-1}{2\alpha} \,N.
\label{Lstat}
\ee
This quantity diverges in the 
thermodynamic limit showing 
the absence of a typical length in such a state. 

For $\alpha <1$, as shown in Appendix~\ref{aless1}, the stationary state behaves as
\be
C_{stat}(r)=\kappa '(N)\,r^{-\alpha}\,
\label{Cstat1}
\ee
with
\be
\kappa '\propto 
N^{-(1-\alpha)}.
\label{depkappa}
\ee
 This is well reproduced by numerical integrations of Eq.~(\ref{eqc2}).

Repeating the calculation as before
one finds again $L_{stat}$ diverging as $N$:
\be
L_{stat}=\frac{1-\alpha}{2(2-\alpha)} \,N.
\label{Lstat1}
\ee

\begin{figure}[ht]
	\vspace{1.0cm}
	\centering
 \rotatebox{0}{\resizebox{0.49\textwidth}{!}{\includegraphics{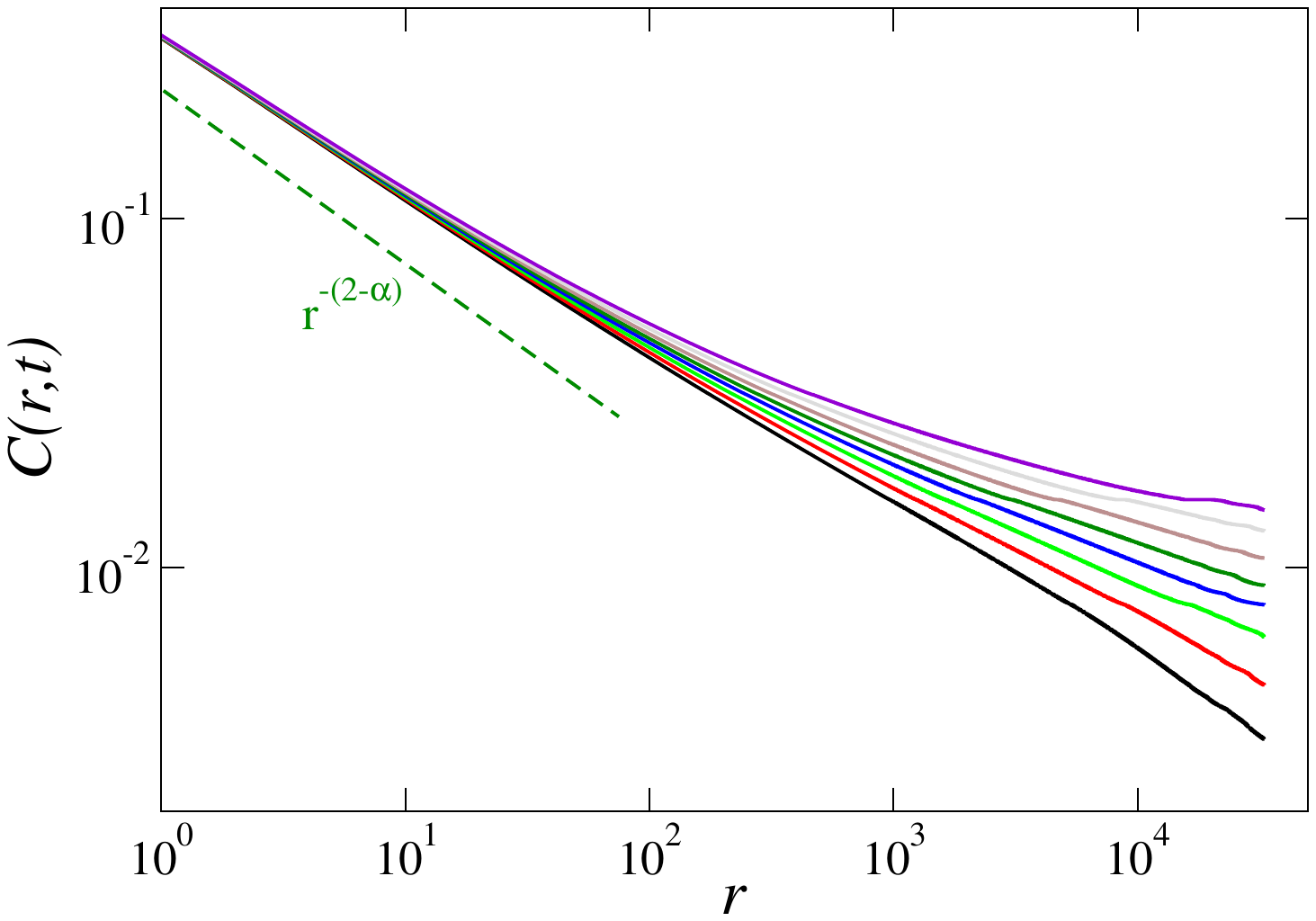}}}
 {\resizebox{0.49\textwidth}{!}
{\includegraphics{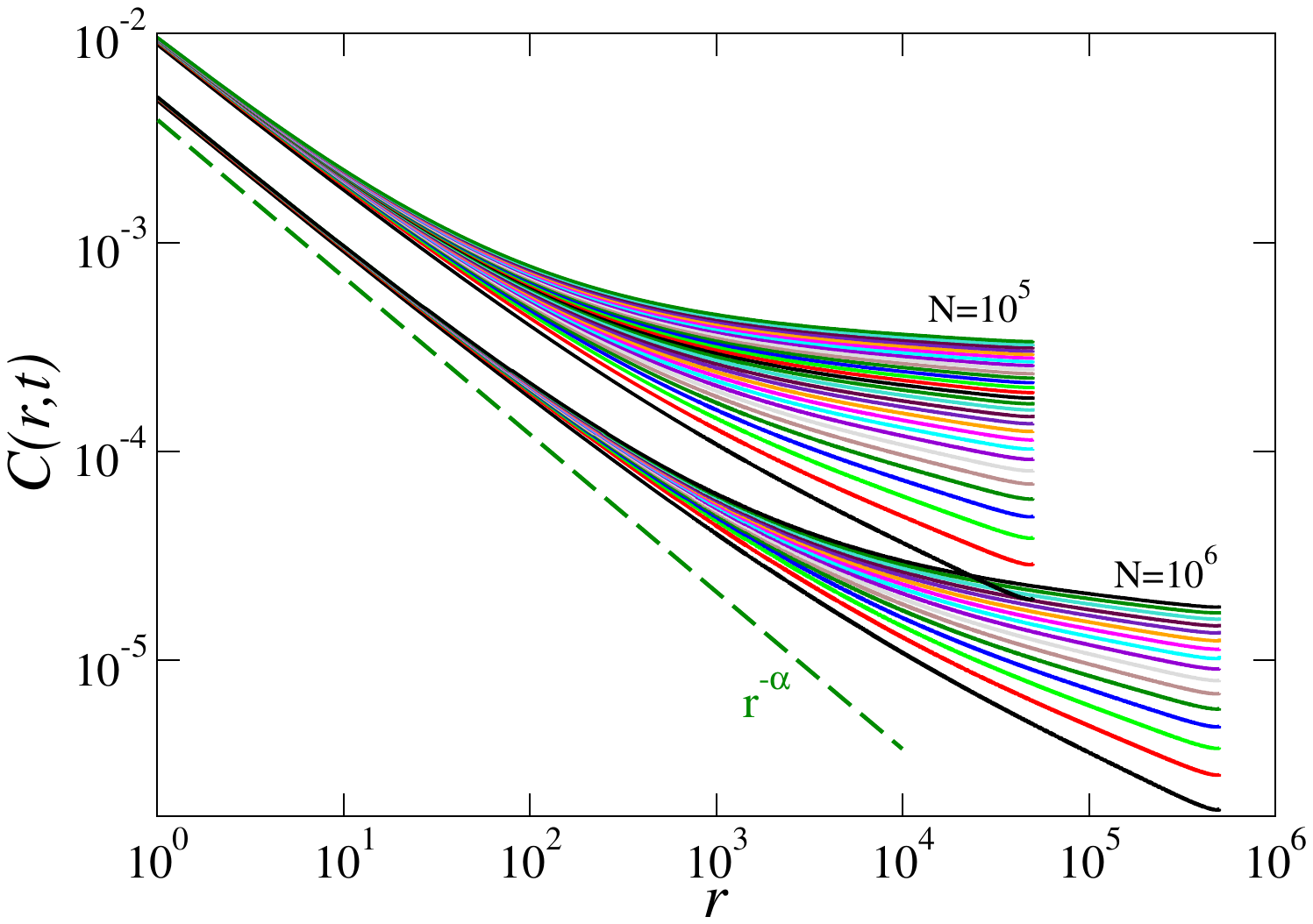}}}
  \caption{$C(r,t)$ is plotted against $r$ for $\alpha=3/2$ (left panel) and $\alpha=3/4$ (right panel). Different curves correspond to different times, increasing from bottom to top (left panel: $t=1300, 1400, \cdots$; right panel: $t=1.5, 2, 2.5, \dots$). Dashed green lines are the analytical behaviors~(\ref{Cstat},~\ref{Cstat1}). On the left panel the system size is $N=10^5$, while the two groups of curves on the right panel correspond $N=10^5$ and $N=10^6$.}
	\label{figCstat}
\end{figure}

In Fig.~\ref{figCstat}, $C(r,t)$ obtained by a numerical solution of Eq.~(\ref{eqc2}) is plotted against $r$ at different times, for $\alpha=3/2$ and for
$\alpha=3/4$. It is seen that $C(r,t)$ is time-independent and obeys the analytic stationary forms~(\ref{Cstat},~\ref{Cstat1}) for sufficiently small values of $r$. Conversely, some evolution takes place at large $r$ which, in a finite system, eventually restores dynamics leading to consensus in a time $T(N)$ that will be determined in Sec.~\ref{ndependence}. However, such incipient kinetics is pushed towards larger and larger values of $r$ as $N$ is increased, as it can be seen in Fig.~\ref{figCstat} for the case $\alpha=3/4$. This is the mechanism whereby a fully stationary state is obtained in the thermodynamic limit.

\subsection{Approaching the stationary state}
\label{approachstat}

As seen in the previous section~\ref{ninftyfirst}, the correlation length $L_{stat}\propto N$ 
in the stationary state is huge. Then, since $L(t)$ starts microscopic initially, it has to grow in the regime preceding stationarity, which happens by means of coarsening.
Such growth is present for any value of $0 \le \alpha\le 2$, however, as we shall see, its duration $t_{coars}$ is extensive in $N$ only for $1< \alpha \le 2$. Then, at least in this range of $\alpha$, this regime is relevant and worth to study, which we do here.

\begin{figure}[ht]
	\vspace{1.0cm}
	\centering
 \rotatebox{0}
 {\resizebox{0.49\textwidth}{!}{\includegraphics{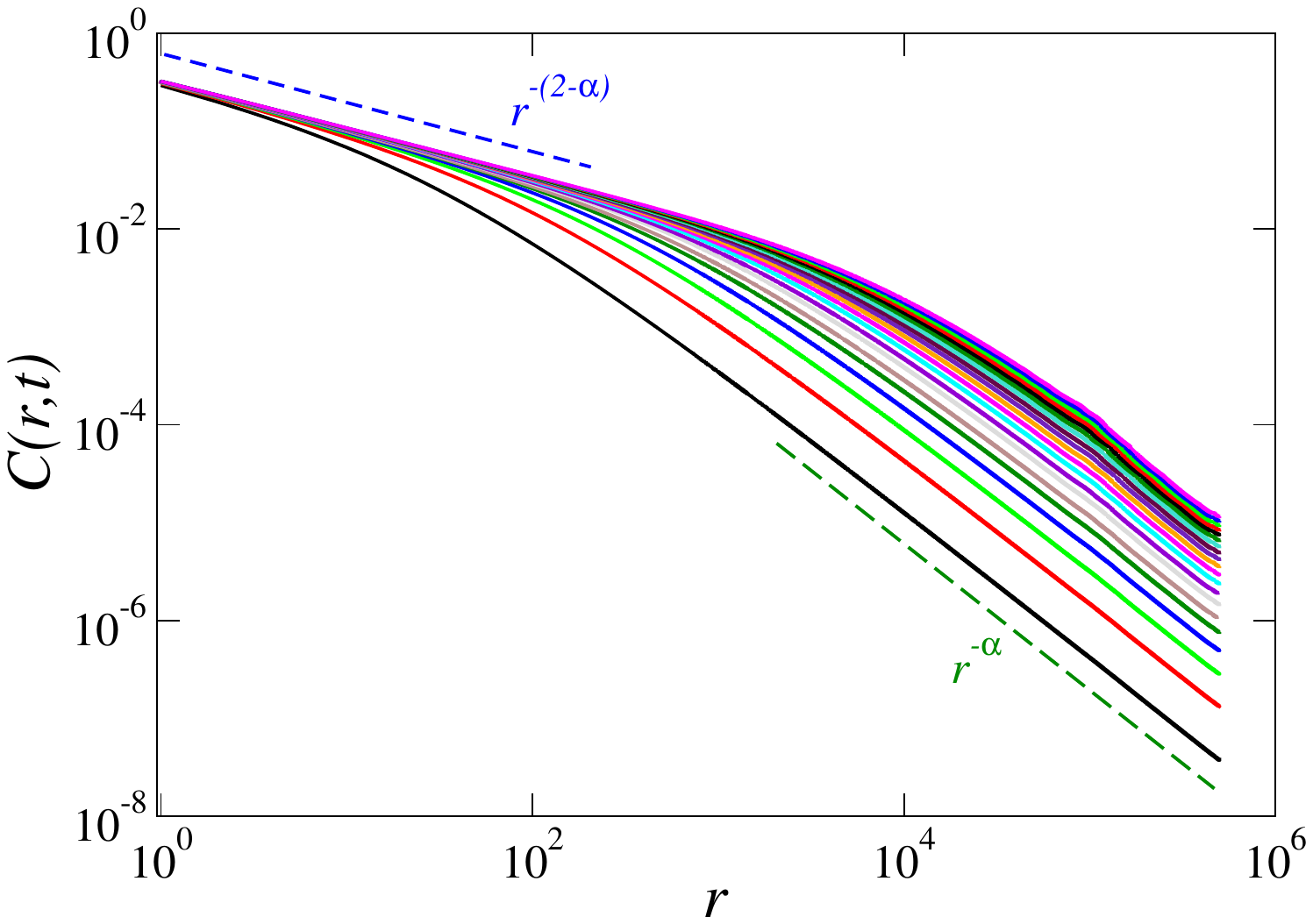}}}
 {\resizebox{0.49\textwidth}{!}{\includegraphics{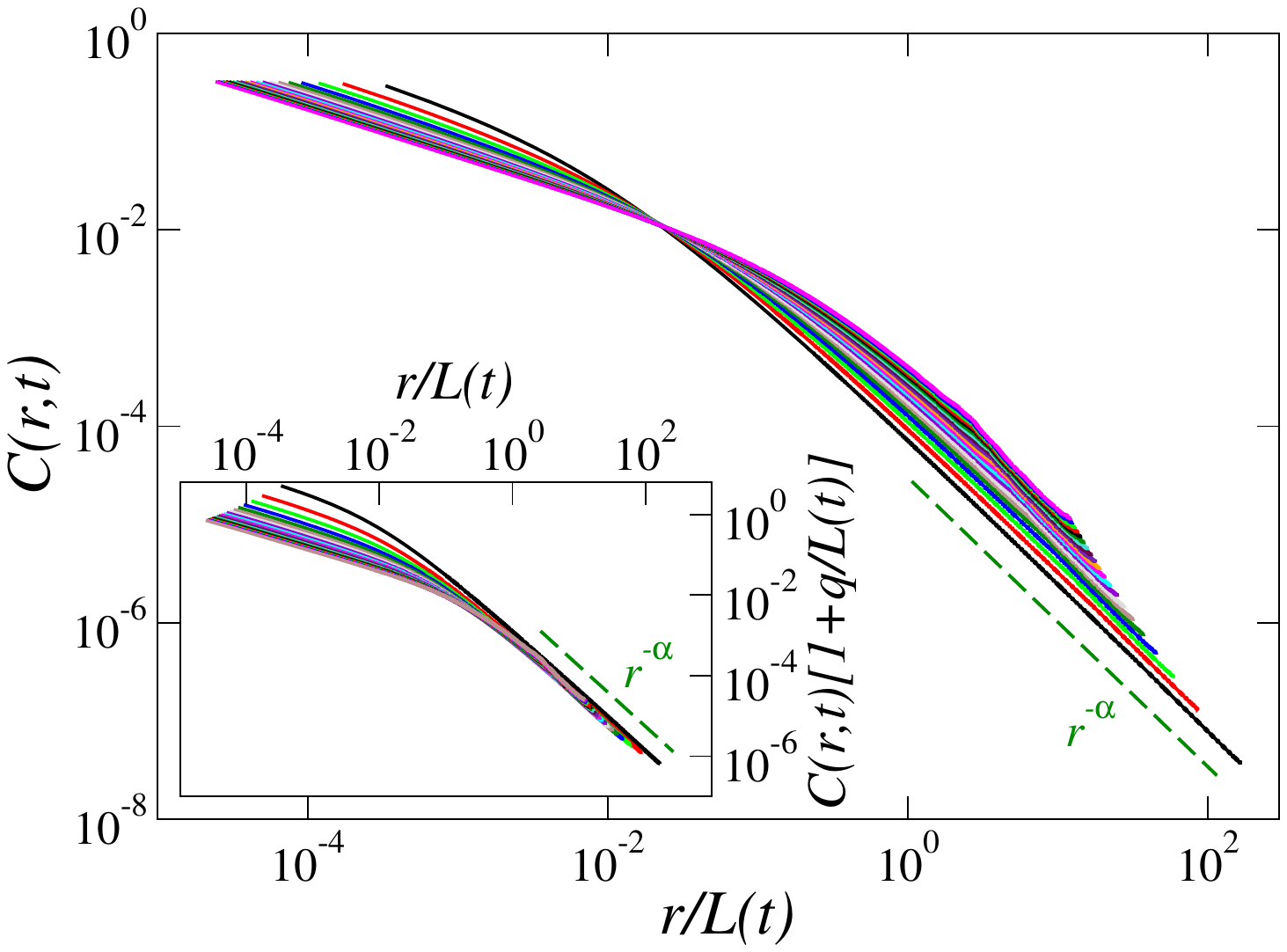}}}
  \caption{$C(r,t)$ is plotted against $r$ (left panel) or against $x=r/L(t)$ (right panel) at different times from $t=5$ to $t=100$ in steps of $\Delta t=5$ (different colors, bottom-top) for $\alpha=3/2$. This range of times corresponds to a ballistic increase of $L(t)$. System size is $N=10^6$. The dashed blue line corresponds to the stationary form~(\ref{Cstat}), while the green line is the behavior~(\ref{unomenxbis}). In the inset we plot $C(r,t)[1+q/L(t)]$, with $q=5\cdot10^4$, against $x$ (same data as main figure, see text for discussion).}
	\label{fig_scal_corr}
\end{figure}

Since the stationary state is highly correlated, it can be expected that such correlations start to develop at small distances, to proceed then to progressively larger values of $r$ as time elapses, until full stationarity.
This is clearly seen in Fig.~\ref{fig_scal_corr},
where it can be appreciated that $C(r,t)$
approaches $C_{stat}$ starting from small values of $r$.

In the coarsening regime we are studying,
the term containing $C(|r-\ell|,t)$
dominates on the r.h.s. of Eq.~(\ref{eqc2}), as in previous cases.
The important contribution in this term occurs at small $|r-\ell|$
where we can use the stationary form~(\ref{Cstat}).  Then it reads
\be
\sum_{\ell}^{0\le r-\ell\le L(t)}P(\ell)\,C_{stat}(|r-\ell|),
\label{term1}
\ee
where, invoking scaling, we have truncated the sum to values of $\ell$ such that $|r-\ell| \le L(t)$ because, after that value, the stationary form 
is abandoned and a faster and integrable 
decay of $C$ is obeyed (see Fig.~\ref{fig_scal_corr}). Such decay, indeed, will be readily shown to be algebraic as $r^{-\alpha}$.
Actually, instead of $|r-\ell| \le L(t)$, in Eq.~(\ref{term1}) we have considered only the region $r-\ell \le L$ because the part with
$\ell-r \le L(t)$ contributes negligibly to the sum because of the decrease of $P(\ell)$. Changing variable to $m=r-\ell$ we have
\be
\sum _{m=0}^{L(t)}P(r-m)\,C_{stat}(m).
\ee
For $r\gg L(t)$, $P(r-m)\simeq P(r)$
so we have
\be
P(r)\sum _{m=0}^{L(t)}C_{stat}(m)=P(r)\left[1+\sum _{m=1}^{L(t)}C_{stat}(m)\right ],
\ee
where we have taken the term with $m=0$ out of the sum (because $C_{stat}(r=0)=1$ and is not described by Eq.~(\ref{Cstat})).
Now we evaluate the sum  using the explicit stationary form~(\ref{Cstat}), finding
a divergence with $L$ as 
\be
kP(r)L(t)^{\alpha -1}\propto L(t)^{-1}x^{-\alpha},
\ee
with $x=r/L(t)$.
Replacing the whole r.h.s. of Eq.~(\ref{eqexp}) with this result 
we get a ballistic increase 
\be
L(t)=A\,t,
\label{balreg}
\ee
of the correlation length and a scaling
function as in~(\ref{unomenxbis}) for large $x$ (for small $x$, as noted already, one has $C(r,t)=C_{stat}(r)$).

The behavior of $C(r,t)$, with its large-$x$ scaling~(\ref{unomenxbis}), can be appreciated in Fig.~\ref{fig_scal_corr}. 
Note that, in the right panel, one sees a poor collapse of the curves, indicating important preasymptotic corrections. Actually, in the inset, it is shown that a tentative correction of the form $C(r,t)[1+q/L(t)]=f(x)$, with $q$ a fitting parameter,
provides a good collapse.

The growth~(\ref{balreg}) can be clearly seen in Fig.~\ref{figL} (cases with $0<\alpha \le 2$). From this figure one sees
a glimpse of ballistic growth also for $\alpha<1$, which however we don't consider analytically here since the duration $t_{coars}$ of such coarsening is microscopic, at variance to the case $1<\alpha \le 2$, where it is of order $N^{\alpha-1}$, as it will be shown at the end of Sec.~\ref{ndependence}.

\section{Size dependence in the coarsening stage} 
\label{ndependence}

In coarsening systems with short-range interactions the finite size of the sample starts to be felt when the typical growing domain size has become comparable to the system size. 
This time is progressively delayed upon increasing the system size; before it, intensive quantities are independent of $N$, for large $N$. 

When interactions are sufficiently long-ranged, due to the one-to-all character of the coupling, there can be a size dependence at any time~\cite{CORBERI2023113681}. 
In the following we show that, in the present model, this happens for $\alpha \le 2$, and we determine such $N$ dependence for the quantity $L(t)$ in the coarsening stage. As a byproduct, this will allow us to estimate, in the next section, the consensus time $T (N)$, i.e., the time taken by the dynamics to reach a fully ordered configuration in
a system of size $N$.

For large $L$,
Eq.~(\ref{eqL}) can be written, by using Eq.~(\ref{scalAnsatzSimple}) 
and transforming the sums into integrals, as
\be
\frac{\int _{\frac{1}{L(t)}}^{\frac{N}{2L (t)}} xf(x)\,dx}{\int _{\frac{1}{L(t)}}^{\frac{N}{2L (t)}}f(x)\,dx}=1.
\label{eqeq1}
\ee
The $N$ dependence comes in due to the upper integration extremes. 
The previous sections showed that, in the coarsening stage, for large $x$ it is $f(x)\sim (ax)^{-\alpha}$ for any value of $\alpha$.
Then, for $\alpha>2$, the integrals converge as the upper integration extremum diverges with $N\to \infty$, which is consistent with
having an $N$-independent integrand, and hence $L$ is independent of $N$.
This can be appreciated in the inset of Fig.~\ref{figL} for the case with $\alpha=2.5$ (blue curve). The comparison between the two sizes $N=10^3$ and $N=10^4$ shows that the 
two curves start superimposed and then separate only when $L(t)$ in the smaller system starts saturating when it becomes of the order of the system size. This is the typical scenario one encounters in a short-range system.

For $1<\alpha\le 2$, after changing variable to $z=ax$ in Eq.~(\ref{eqeq1}), leading to
\be
\frac{1}{a}\,\frac{\int _{\frac{a}{L(t)}}^{\frac{aN}{2L (t)}} zf(z)\,dz}{\int _{\frac{a}{L(t)}}^{\frac{aN}{2L (t)}}f(z)\,dz}=1,
\label{eqeqeq1}
\ee
one sees that the integral in the numerator diverges as $\left [\frac{aN}{L(t)}\right ] ^{2-\alpha}$.  
Let us note now that, because of the form~(\ref{unomenxbis}) of the scaling function,
$L$ and $a$ must share the same $N$-dependence.
Indeed, if this were not true, in the thermodynamic limit $N\to \infty$
 one would have $C\equiv 1$ or $C\equiv 0$ for any $r>0$ at finite times, which is unphysical.
Using this fact,
 one has a $N^{2-\alpha}$ divergence in the numerator of Eq.~(\ref{eqeqeq1}).
The denominator, instead, is constant (in $N$). 
Then, in order for the $N$ dependence to drop out from the l.h.s. of Eq.~(\ref{eqeqeq1}) it must be
$L(t)\propto N^{2-\alpha}$.
As discussed at the beginning of this section, this size dependence is different from the one usually displayed by short-range systems, because it occurs at any time
$t>0$, as it can be appreciated in the inset of Fig.~\ref{figL} in the case $\alpha=1.8$.

For $\alpha \le 1$ both the integrals in the numerator and in the denominator of Eq.~(\ref{eqeqeq1}) diverge with $N$ at the upper extreme so that, proceeding as before, the l.h.s. turns out to behave as $\frac{N}{L(t)}$, leading to $L(t)\propto N$.

In summary
\be
L(t)\propto \left \{ \begin{array}{ll}
N^0, &\quad \quad \alpha >2, \\
\ln N, &\quad \quad \alpha =2, \\
N^{2-\alpha}, &\quad \quad 1<\alpha < 2, \\
N/\ln N, &\quad \quad \alpha=1, \\
N, &\quad \quad \alpha < 1,
\end{array} \right . 
\label{Nbeh}
\ee
where we have made explicit the logarithmic corrections occurring for the boundary values $\alpha=1$ and $\alpha=2$. 
Note the unusual fact of $L$ growing unboundedly in the thermodynamic limit $N\to \infty$
even at finite times for $\alpha \le 2$.
Furthermore, for $\alpha <1$, $L$ reaches the system size in a time of order $N^0$ and hence
there is not a real coarsening stage extending in time as $N$ gets large.

We have checked that all these results are very well reproduced by a numerical integration of Eq.~(\ref{eqc2}). 

To conclude this section, let us compute the duration of the coarsening stage for $\alpha \le 2$. Using Eq.~(\ref{balreg}),
with $A\propto N$ according to Eq.~(\ref{Nbeh}), stipulating that coarsening ends when $L(t)\simeq L_{stat}$ one readily obtains the 
coarsening duration
\be
t_{coars}\propto N^{\alpha-1}
\ee
for $1<\alpha \le 2$. This result shows that
coarsening is a macroscopic phenomenon for $\alpha >1$.

\section{Consensus time} \label{secconsensus}

In the following we discuss the time $T(N)$ taken by the model to reach consensus.
For $\alpha>2$, when ordered domains are well defined and coarsen, $T(N)$ can be 
obtained from the same condition $L(t=T) \simeq N$ as for short range 
interactions, namely consensus is achieved when the typical size of correlated regions reaches the system size.
For $\alpha<2$, instead, consensus is not reached although $L \simeq N$, because spins inside 
ordered regions can frequently flip by interacting with very far ones. As a consequence, the reaching of the consensus occurs via a global symmetry
breaking as in mean-field and the consensus time is proportional to $N$. 
In summary:  
\be
T(N) \propto \left \{
\begin{array}{ll}
N^2, &\quad \quad \alpha >3, \\
N^{\alpha-1}, &\quad \quad 2<\alpha \le 3, \\
N/\ln N, &\quad \quad \alpha=2, \\
N, &\quad \quad \alpha <2,
\end{array} \right .
\label{Nbeh2}
\ee

\begin{figure}[ht]
	\vspace{1.0cm}
	\centering
	\rotatebox{0}{\resizebox{0.5\textwidth}{!}{\includegraphics{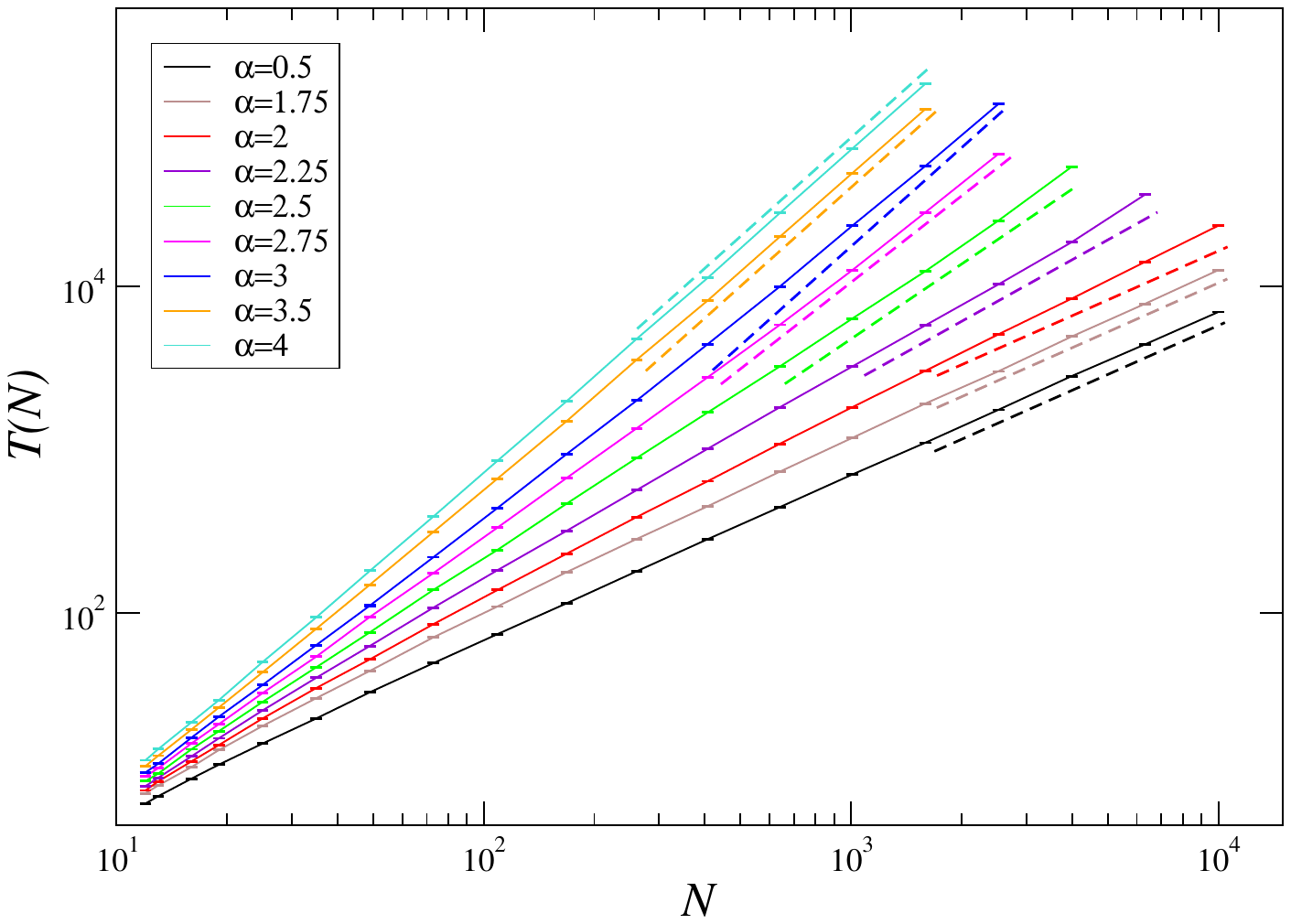}}} 
  \caption{The consensus time (in Monte Carlo step units), computed by means of numerical simulations of the model, is plotted against the system size $N$ for different
  values of $\alpha$, see legend. Every curve is an average over $10^4$ realisations. Data points are represented with their (very small) error bars. Continuous lines are guides for the eye. The dashed lines near each dataset (same color) are the expected behaviors~(\ref{Nbeh2}). 
  }
	\label{fig_consensus}
\end{figure}

The consensus time $T(N)$ is a suitable quantity to check the validity of our approach by comparing analytical predictions~(\ref{Nbeh2})
with numerical simulations, i.e., evolving the spins according to a standard Monte Carlo procedure with the transition rates~(\ref{transrates}). Results in  Fig.~\ref{fig_consensus} indicate a very good agreement, which confirms the correctness of the whole analytical approach. 

\section{Conclusions}\label{concl}

In this paper we have studied analytically the ordering kinetics of the one-dimensional voter model with long-range interactions.
This provides clear conclusions that are very difficult to extract from numerical simulations alone,
particularly when the decay of interactions with distance becomes slow, because of scaling violations, massive finite-size effects
and strong noise. Indeed, the same problem was considered 
in~\cite{PhysRevE.83.011110}, where the authors, based solely on numerical simulations, arrived at partially different conclusions. 
For instance, they find that the exponent regulating the time-decay of the interface density 
attains the NN value for $\alpha \ge 6$, whereas we find that this happens for $\alpha \ge 3$, see Eq.~(\ref{expdecayrhot}).

Our solution uncovers a rather rich pattern of behaviors
as $\alpha $ is varied. With respect to the standard paradigm of phase-ordering in scalar systems with short-range interactions, 
at least two new features emerge due to the effect of long-range interactions. 
The first is related to violations
of dynamical scaling in its simplest formulation. Indeed, already for $\alpha >3$, where the growth exponent remains unchanged with respect to the NN case, some sizeable scaling corrections, which are absent for nearest-neighbours interactions, affect the large-distance behavior, although their effect is slowly suppressed as time elapses. 
Violations become more severe and unescapable for $\alpha <3$, with the simultaneous presence of two ordering lengthscales,
behaving differently. The first, $\rho^{-1}$, is associated to the size of growing domains and the other, $L$, corresponds to the increasing range of correlations.
The second new feature is the presence of long-lived non-equilibrium stationary states, 
which are a hallmark of systems with long-range
interactions but were never observed before, to the best of our knowledge, in long-range classical statistical mechanical out of equilibrium spin systems.
In the present model such states are present for $\alpha <2$ and their lifetime diverges in the thermodynamic limit.

Let us comment on the similarities and the differences between the behavior of the present model and of the Ising model with Glauber dynamics (still in $d=1$, with a coupling constant
$J(r)\propto r^{-\alpha}$ between spins) after a quench to zero temperature. 
For NN interactions (i.e. $\alpha \to \infty$)
the two systems are equivalent. For the algebraic interactions considered here, they still 
behave similarly for $\alpha >\alpha _{SR}$, where $\alpha _{SR}$ is a critical value above which ordered regions 
grow as $t^{1/2}$, as in the NN case. For the voter model we have shown that $\alpha _{SR}=3$, while it is known that in the Ising case it is $\alpha_{SR}=2$~\cite{CLP_review, PhysRevE.49.R27, PhysRevE.50.1900}. For both models, there exist another characteristic value $\alpha _{LR}$, below which a mean-field phenomenology starts to be observed.
By mean-field phenomenology we mean, in a broad sense, that a true coarsening behavior, akin to the one observed for short-range interactions, is
not displayed. In the voter model it is $\alpha _{LR}=2$ and the mean-field phenomenology amounts to the presence of metastable stationary states becoming infinitely long-lived and preventing consensus to be reached in the thermodynamic limit $N\to \infty$. In the Ising model there are no such states and the mean-field 
phenomenology is manifested by the fact that, for $\alpha <\alpha_{LR}=1$, the statistical ensemble is dominated by histories where ordering is 
achieved by aligning all the spins with the sign of the majority in a time of order one, without formation of domains, as it happens in the fully mean-field case
($\alpha =0$). For intermediate $\alpha$ values, namely for $\alpha_{LR}<\alpha <\alpha_{SR}$, there is coarsening with an $\alpha$-dependent exponent, 
$L(t)\propto t^{1/(1+\alpha-\alpha_{LR})}$. This is true for both models (with the different values of $\alpha_{SR}$ and $\alpha_{LR}$ mentioned above)
with the important difference that in the voter model scaling is violated due to the simultaneous presence of the extra length $\rho^{-1}(t)$ next to $L(t)$,
see Sec.~\ref{aless2}. Note also that the characteristic values $\alpha_{SR}$ and $\alpha_{LR}$ are larger by one unit for the voter model with respect to the Ising one. 
This can be understood as due to the fact that in the latter the interaction is mediated by the Weiss local field, which is a quantity integrated over an extensive region, while in the former the interaction is one to one. Despite the differences, the two models exhibit some common features; we hope that the solution presented here of the voter model can trigger a better understanding of the kinetics of the
Ising model with extended interactions, for which a a closed equation for the correlators cannot be written. As a very final remark, let us mention that such a closed equation can be obtained for other quantities besides the ones studied here, e.g. two-time quantities~\cite{CorSmalTwoTime}, and in any dimension~\cite{PhysRevE.109.034133}.

\section*{Acknowledgements}

We thank Luca Smaldone for discussions. F.C. acknowledges financial support by MUR PRIN PNRR 2022B3WCFS.

\appendix

\section{Density of interfaces} \label{appdensint}
	
Indicating with $C_1(t)$ the correlation $C(r=1,t)$ at unitary distance, for this quantity 
Eq.~(\ref{eqc2}) reads 
\be
\dot C_1(t)=2[2{\cal S}(t)-C_1(t)],
\label{C1}
\ee
where 
\be
2{\cal S}(t)=\sum _{\ell=1}^{N/2}P(\ell)C(\ell,t)
\label{cals}
\ee
and we have performed the approximation
$C([[1-\ell]],t)+C([[1+\ell]],t)\simeq 2C(\ell,t)$.
Let us assume
\be
C(\ell,t)=e^{-\left (\frac{\ell}{L(t)}\right )^\beta},
\ee
where $\beta$ is a parameter to be determined. 
Plugging this form into Eq.~(\ref{cals}), assuming $0<\beta<1$ 
and extending the sums 
to infinity, after taking the leading order in $1/L(t)$ one arrives at
\be
2{\cal S}(t)=1-\frac{1+2\beta+\alpha^2-2\alpha(1+\beta)}{(\alpha+1-2\beta)(\alpha-1-\beta)}\,L(t)^{-\beta}.
\label{2S}
\ee
Recalling that the density of interfaces (anti-aligned adjacent spins)
$\rho$ is related to $C_1$ by $C_1=1-2\rho$ and inserting
Eq.~\eqref{2S} into Eq.~(\ref{C1}), we obtain
\be
\dot \rho = -2\rho + k' L^{-\beta},
\ee
where $k'$ is the prefactor of $L(t)^{-\beta}$ in Eq.~\eqref{2S}.
Since $\beta<1$,  $\dot \rho$ is negligible with respect to both terms on the r.h.s. and we arrive at
\be
\rho(t) \sim\frac{k'}{2} L(t)^{-\beta}.
\ee	

\section{The exponent \boldmath{$\beta$}} \label{appbeta}

In this Appendix we validate numerically the scaling arguments presented in Appendix~\ref{appdensint} assuming that the relation between $\beta$ and $\alpha$ is
$\beta=\alpha-2$.
We solve numerically Eq.~\eqref{eqc2} for the correlation function
for a system of $N$ sites and from $C(r,t)$ we determine the
evolution of $1-2\cal{S}$ (using Eq.~\eqref{cals}), of $\rho(t)$ (using
Eq.~\eqref{defrho}) and of $L(t)$ (using Eq.~\eqref{eqL}).
In the figures we plot the effective exponents fitted locally around
a given time (or $L$) and compare them to the values predicted assuming
$\beta=\alpha-2$.
The interpretation of the results is made difficult by the presence of
large preasymptotic temporal regimes combined with strong
finite size effects. Yet the numerical evidence is sufficient to 
strongly support the validity of the theoretical analysis.

 \begin{figure}[ht]
	\vspace{1.0cm}
	\centering
	\rotatebox{0}{\resizebox{0.5\textwidth}{!}{\includegraphics{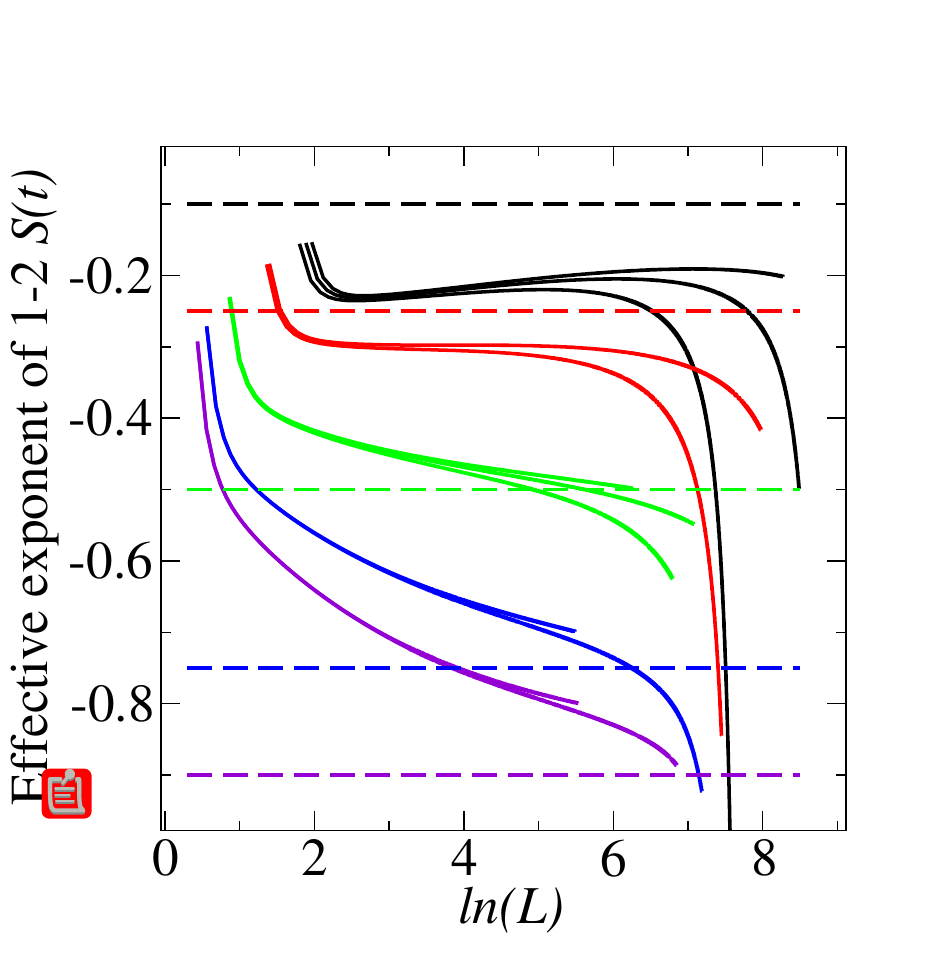}}} 
  \caption{Effective exponent of $1-2{\cal S}$ as a function of $\ln L$
  for $\alpha=2.1$ (black), $\alpha=2.25$ (red), $\alpha=2.5$ (green), $\alpha=2.75$ (blue), $\alpha=2.9$ (violet), compared with the prediction $-(\alpha-2)$ (horizontal dashed lines). Different solid lines, from bottom to top, are for $N=10^4$, $N=3\cdot 10^4$, $N=10^5$.}
	\label{fig_EffExp1-2S}
\end{figure}

In Fig.~\ref{fig_EffExp1-2S} we analyze the evolution of the effective exponent of
$1-2\cal{S}$ as a function of $\ln L$. 
Although in a transient regime it exhibits a non-monotonic behavior,
for all values of $\alpha$ it turns out that for long times, before finite size
effects show up (manifested by a fast decrease of the curves), the effective exponents tends toward the expected value.

 \begin{figure}[ht]
	\vspace{1.0cm}
	\centering
	\rotatebox{0}{\resizebox{0.5\textwidth}{!}{\includegraphics{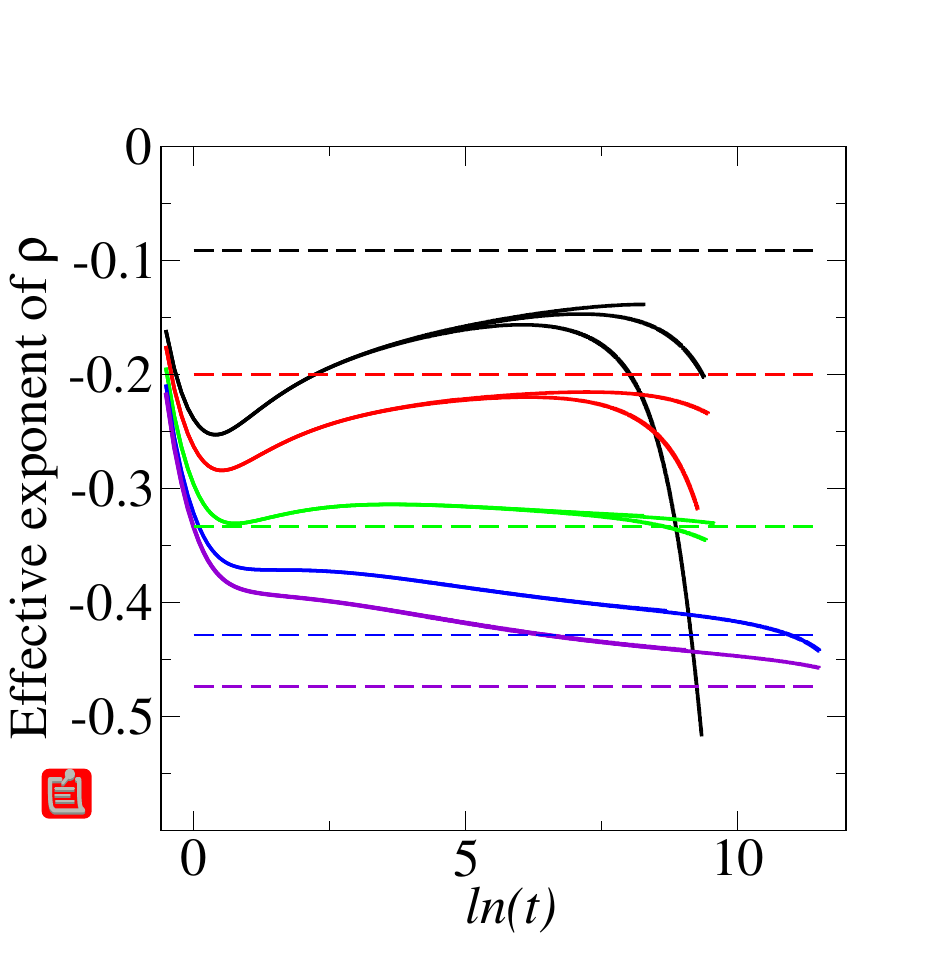}}} 
  \caption{Effective exponent of $\rho$ as a function of $\ln t$
  for $\alpha=2.1$ (black), $\alpha=2.25$ (red), $\alpha=2.5$ (green), $\alpha=2.75$ (blue), $\alpha=2.9$ (violet), compared with the prediction $-(\alpha-2)/(\alpha-1)$ (horizontal dashed lines). Different solid lines, from bottom to top, are for $N=10^4$, $N=3\cdot 10^4$, $N=10^5$.}
	\label{fig_EffExprho}
\end{figure}

\begin{figure}[ht]
	\vspace{1.0cm}
	\centering
	\rotatebox{0}{\resizebox{0.5\textwidth}{!}{\includegraphics{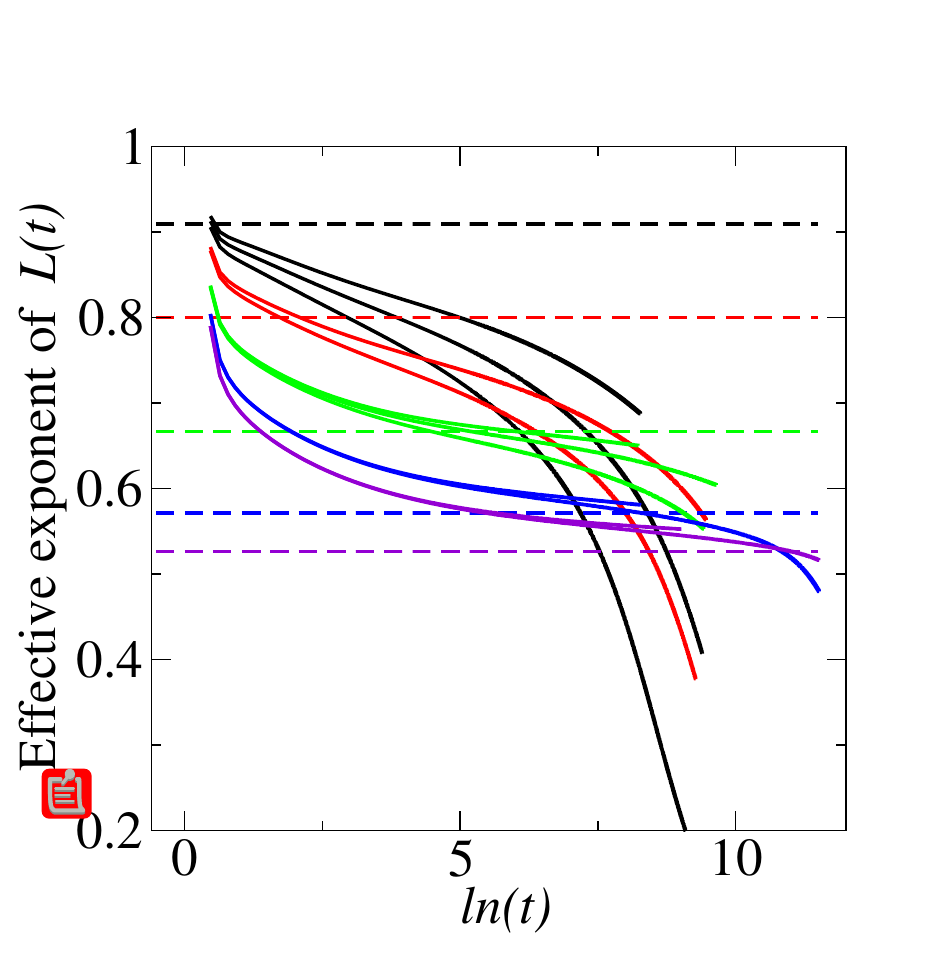}}} 
  \caption{Effective exponent of $L$ as a function of $\ln t$
  for $\alpha=2.1$ (black), $\alpha=2.25$ (red), $\alpha=2.5$ (green), $\alpha=2.75$ (blue), $\alpha=2.9$ (violet), compared with the prediction $1/(\alpha-1)$ (horizontal dashed lines). Different solid lines, from bottom to top, are for $N=10^4$, $N=3\cdot 10^4$, $N=10^5$.}
	\label{fig_EffExpL}
\end{figure}

The same conclusion can be drawn when considering the evolution of the effective exponent of $\rho$ vs $\ln t$ (Fig.~\ref{fig_EffExprho}) and the effective exponent of
$L$ vs $\ln t$ (Fig.~\ref{fig_EffExpL}), although in this last case the preasymptotic corrections are even stronger.

\section{Stationary states}
\subsection{Case \boldmath{$1 < \alpha \le 2$}}\label{ain12}

Setting the l.h.s. of Eq.~(\ref{eqc2}) to zero and using the form~(\ref{Cstat}) in the sums, one has
\begin{eqnarray}
	\frac{Z}{\kappa}\,C_{stat}(r)&=& \sum_{\ell=1}^{r-1}\ell^{-\alpha}(r-\ell)^{-(2-\alpha)}+r^{-\alpha}+\sum_{\ell=r+1}^{N/2}\ell^{-\alpha}(\ell-r)^{-(2-\alpha)}+
	\sum_{\ell=1}^{N/2-r}\ell^{-\alpha}(r+\ell)^{-(2-\alpha)}+\nonumber \\
	&+&\sum_{\ell=N/2-r+1}^{N/2}\ell^{-\alpha}(N-r-\ell)^{-(2-\alpha)}.
	\label{statsums}
\end{eqnarray}
Transforming the sums into integrals one obtains 
\begin{eqnarray}
	\frac{Z}{\kappa}\,C_{stat}(r)&=&-\frac{(r-1)^{-(\alpha -1)}}{(\alpha -1)r}+\frac{(r-1)^{\alpha -1}}{(\alpha -1)r}+r^{-\alpha}+
	\frac{\frac{N}{2}^{-(\alpha -1)}(\frac{N}{2}-r)^{\alpha -1}}{(\alpha -1)r}-\frac{(r+1)^{-(\alpha -1)}}{(\alpha -1)r}
	-\frac{(\frac{N}{2}-r)^{-(\alpha -1)}\frac{N}{2}^{\alpha -1}}{(\alpha -1)r}+\nonumber \\
	&+&\frac{(r+1)^{\alpha -1}}{(\alpha -1)r}
	-\frac{\frac{N}{2}^{-(\alpha -1)}(\frac{N}{2}-r)^{\alpha -1}}{(\alpha -1)(N-r)}+\frac{(\frac{N}{2}-r+1)^{-(\alpha -1)}(\frac{N}{2}-1)^{\alpha -1}}{(\alpha -1)(N-r)}.
\end{eqnarray}
For $r\gg1$ and $\frac{r}{N}\ll 1$ the dominant contributions in the r.h.s. are the second and the seventh term, leading to
\be
\frac{Z}{\kappa}\,C_{stat}(r)\simeq \frac{2}{\alpha -1}\,r^{-(2-\alpha)},
\ee
which is consistent with our original ansatz~(\ref{Cstat}).

\subsection{Case \boldmath{$\alpha <1$}}\label{aless1}

Plugging the ansatz~(\ref{Cstat1}) into Eq.~(\ref{eqc2}), and letting $\frac{d}{dt} C=0$, one obtains 
\begin{eqnarray}
	Zr^{-\alpha}&=& \sum_{\ell=1}^{r-1}\ell^{-\alpha}(r-\ell)^{-\alpha}+r^{-\alpha}+\sum_{\ell=r+1}^{N/2}\ell^{-\alpha}(\ell-r)^{-\alpha}+\nonumber \\
	&+&\sum_{\ell=1}^{N/2-r}\ell^{-\alpha}(\ell+r)^{-\alpha}
	+\sum_{\ell=N/2-r+1}^{N/2}\ell^{-\alpha}(N-r-\ell)^{-\alpha}.
	\label{statsums1}
\end{eqnarray}
For small $r$, this expression is dominated by the contribution provided by the terms in 
the first line. There, the sums have a prominent peak around $\ell=r$. So we can evaluate them by letting $\ell ^{-\alpha}\simeq r^{-\alpha}$ and transforming  
the sums into integrals. We get
\be
Zr^{-\alpha}\simeq \frac{(N/2)^{1-\alpha}}{1-\alpha}\, r^{-\alpha},
\ee
which is consistent.
For large $r$, $r\lesssim N/2$, the first sum prevails, which can be 
written $\sum _\ell \ell^{-\alpha} r^{-\alpha}$, because $r\gg \ell$ almost everywhere, providing again consistency.

Now we have to determine $\kappa '$. Let us stipulate that the form~(\ref{Cstat1}) holds up to a certain $r^*=\gamma N$ of order $N$ ($\gamma <1$ is a constant).
One has $\sum _{r=1}^{r^*}C_{stat}(r)=\langle S_i \sum_{j=i+1}^{i+r^*}S_j\rangle =\gamma N\langle S_i\rangle m=\gamma N m^2$, where $m=\langle S_i\rangle$ is the magnetisation and we have assumed self-averaging, i.e. $\sum_{j=i+1}^{i+r^*}S_j=\gamma N m$. Since the stationary state is disordered, one must have $m\propto N^{-1/2}$, due to the central limit theorem. Hence $\sum _{r=1}^{r^*}C_{stat}(r)\propto N^0$ which, with the form~(\ref{Cstat1}), implies $\kappa '$ as in Eq.~(\ref{depkappa}).

\bibliography{voterbib}

\end{document}